\documentclass{iau} 
\usepackage{graphicx}

\title[Hot subdwarfs in the Galactic halo] %% give here short title %%
{Hot subdwarf stars in the Galactic halo \\ Tracers of prominent events in late stellar evolution}

\author[Stephan Geier et al.]   %% give here short author list %%
{Stephan Geier$^{1,2}$, Thomas Kupfer$^3$, Veronika Schaffenroth$^{2,4}$,
 Ulrich Heber$^4$ \and the MUCHFUSS collaboration}

\affiliation{$^1$European Southern Observatory, Karl-Schwarzschild-Str.~2, 85748 Garching, Germany \\[\affilskip]
$^2$Dr. Karl Remeis-Observatory \& ECAP, Astronomical Institute, Friedrich-Alexander University Erlangen-N\"urnberg, Sternwartstr. 7, D~96049 Bamberg, Germany \\[\affilskip] 
$^3$Department of Astrophysics/IMAPP, Radboud University Nijmegen, P.O. Box 9010, 6500 GL Nijmegen, The Netherlands \\[\affilskip]
$^4$Institute for Astro- and Particle Physics, University of Innsbruck, Technikerstr. 25/8, 6020 Innsbruck, Austria}

\pubyear{2015}
\volume{317}  %% insert here IAU Symposium No.
\jname{The General Assembly of Galaxy Halos: Structure, Origin and Evolution}
\editors{A. Bragaglia, M. Arnaboldi, M. Rejkuba \& D. Romano, eds.}
\begin{document}

\maketitle

\begin{abstract}
Hot subdwarf stars (sdO/Bs) are the stripped cores of red giants
located at the bluest extension of the horizontal branch. They constitute the
dominant population of UV-bright stars in old stellar environments and are most
likely formed by binary interactions. We perform the first systematic,
spectroscopic analysis of a sample of those stars in the Galactic halo based on
data from SDSS. In the course of this project we discovered 177 close
binary candidates. A significant fraction of the sdB binaries turned out to 
have close substellar companions, which shows that brown dwarfs and planets can 
significantly influence late stellar evolution. Close hot subdwarf binaries 
with massive white dwarf companions on the other hand are good candidates 
for the progenitors of type Ia supernovae. We discovered a hypervelocity star, 
which not only turned out to be the fastest unbound star known in our Galaxy, but 
also the surviving companion of such a supernova explosion. 
\end{abstract}


Hot subdwarf stars (sdO/Bs) are evolved core helium-burning stars with very thin hydrogen envelopes. About half of the sdB stars are in close binaries and are formed by common envelope ejection. The companions are in most cases either late main sequence  stars of spectral type M or compact objects like white dwarfs (WDs). Subdwarf binaries with massive CO-WD companions are candidates for supernova type Ia (SN~Ia) progenitors. The project Massive Unseen Companions to Hot Faint Underluminous Stars from SDSS (MUCHFUSS) aims at finding the sdB binaries with the most massive compact companions like massive WDs, neutron stars or black holes as well as the least massive companions like substellar objects (e.g. Geier et al. \cite{geier15b}). 
                    
We selected and classified about $\sim1400$ hot subdwarf stars from the Sloan Digital Sky Survey (SDSS DR7) by colour selection and visual inspection of their spectra. Stars with high velocities have been reobserved and individual SDSS spectra have been analysed. In total $177$ radial velocity variable subdwarfs have been dis\-covered and $1914$ individual radial velocities measured. We constrain the fraction of close massive companions of H-rich hot subdwarfs to be smaller than $\sim1.3\%$ (Geier et al. \cite{geier15b}). Light curves with a duration of typically 2-3 hours have been obtained of 66 subdwarf binaries from our target list. We found three eclipsing systems, two of them with the first confirmed brown dwarf companions, and one hybrid sdB pulsator with reflection effect.

\begin{figure}[t]
% \vspace*{-2.0 cm}
\begin{center}
 \includegraphics[width=6.5cm]{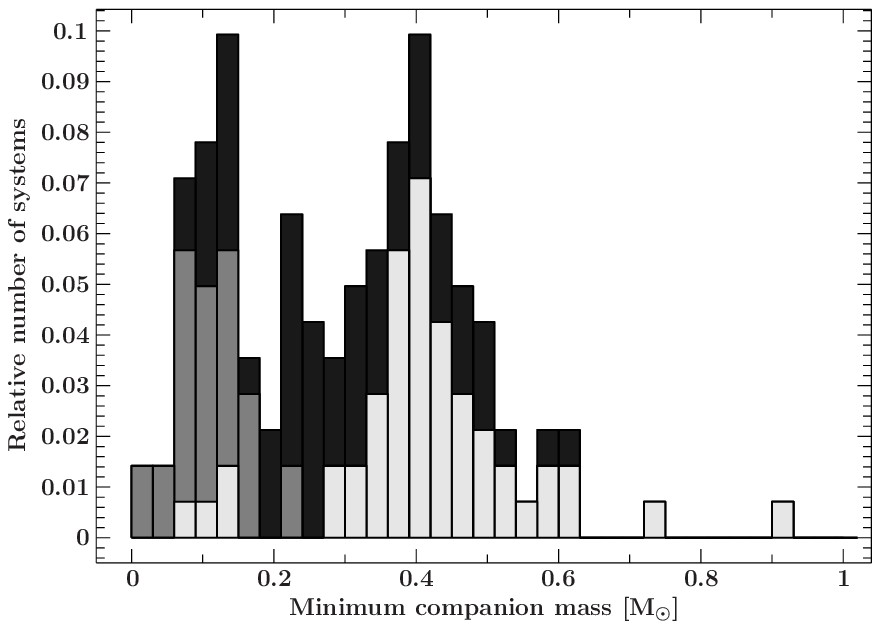}
 \includegraphics[width=6.5cm]{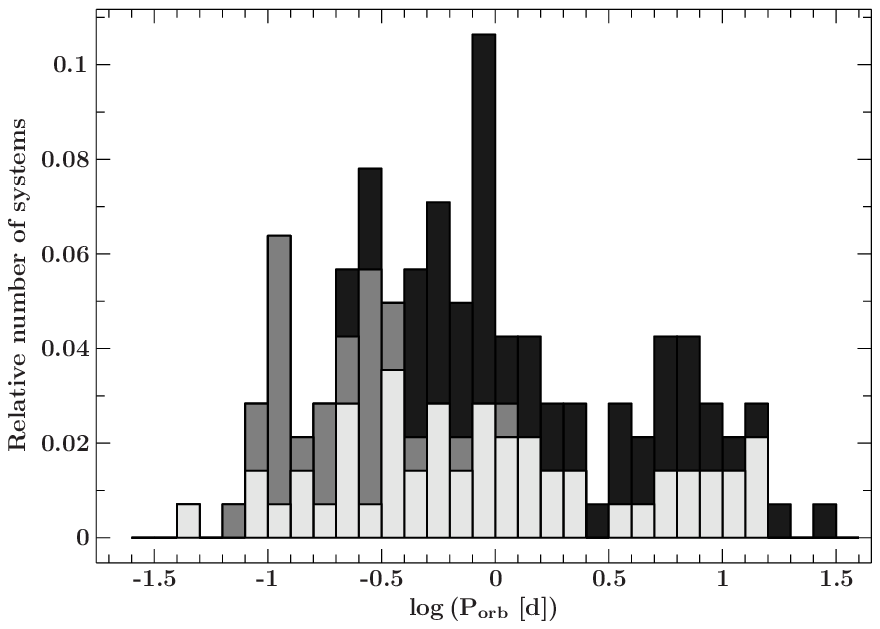}
% \vspace*{-1.0 cm}
 \caption{Histogram of minimum companion masses and orbital periods (light grey: WD companions, grey: dM companion, dark grey: unknown companion type).}
   \label{fig1}
\end{center}
\end{figure}

We determined orbital parameters of 30 close sdB binaries and studied the known population. The distribution of minimum companion masses is bimodal. One peak around $0.1\,M_{\rm \odot}$ corresponds to the low-mass main sequence (dM) and substellar companions. The other peak around $0.4\,M_{\rm \odot}$ corresponds to the WD companions. The derived masses for the WD companions are significantly lower than the average mass for single carbon-oxygen WDs (See Fig.~\ref{fig1}, Kupfer et al. \cite{kupfer15}).

In the course of the MUCHFUSS project, we discovered the first close sdB systems with brown dwarf companions and, subsequently, another even closer system (Geier et al. \cite{geier11}; Schaffenroth et al. \cite{schaffenroth14,schaffenroth15}). Such binaries are important to probe common envelope evolution and study the yet unknown influence of substellar objects like brown dwarfs or giant planets on stellar evolution. We derive a number fraction of $>4\%$ close sdB binaries with brown dwarf companions and a similar fraction of close binaries with M dwarf companions. Substellar companions are therefore as frequent as low-mass stellar companions at short orbital periods.

We detected high RV-variability of the bright sdB CD$-$30$^\circ$11223. Photometric follow-up revealed both shallow transits and eclipses. The binary system, which is composed of a carbon/oxygen WD ($\sim0.76\,M_{\rm \odot}$) and an sdB ($\sim0.51\,M_{\rm \odot}$), has the shortest orbital period ($P\simeq0.049\,{\rm d}$) ever measured for a hot subdwarf binary (Geier et al. \cite{geier13}). In the future the system will transfer mass from the helium star to the WD. After a critical amount of helium is deposited on the surface, the helium is ignited. Modelling this process shows that the detonation in the accreted helium layer is sufficiently strong to trigger the explosion of the core. Thermonuclear supernovae have been proposed to originate from this so-called double-detonation of a WD. The surviving donor hot subdwarf star will then be ejected with its orbital velocity. The properties of such a remnant match the hypervelocity star US\,708, a He-sdO moving with $\sim1200\,{\rm km\,s^{-1}}$. This object is the fastest unbound star known in our Galaxy (Geier et al. \cite{geier15a}).

\end{document}